\documentstyle[12pt]{article}
\def\singlespace {\smallskipamount=3.75pt plus1pt minus1pt
                  \medskipamount=7.5pt plus2pt minus2pt
                  \bigskipamount=15pt plus4pt minus4pt
                  \normalbaselineskip=15pt plus0pt minus0pt
                  \normallineskip=1pt
                  \normallineskiplimit=0pt
                  \jot=3.75pt
                  {\def\smallskip {\vskip\smallskipamount}}
                  {\def\medskip   {\vskip\medskipamount}}
                  {\def\bigskip   {\vskip\bigskipamount}}
                  {\setbox\strutbox=\hbox{\vrule
                    height10.5pt depth4.5pt width 0pt}}
                  \parskip 4.5pt
                  \normalbaselines}
\def\middlespace {\smallskipamount=5.825pt plus1.5pt minus1.5pt
                  \medskipamount=11.25pt plus3pt minus3pt
                  \bigskipamount=22.5pt plus6pt minus6pt
                  \normalbaselineskip=22.5pt plus0pt minus0pt
                  \normallineskip=1pt
                  \normallineskiplimit=0pt
                  \jot=5.825pt
                  {\def\smallskip {\vskip\smallskipamount}}
                  {\def\medskip   {\vskip\medskipamount}}
                  {\def\bigskip   {\vskip\bigskipamount}}
                  {\setbox\strutbox=\hbox{\vrule
                    height15.75pt depth6.75pt width 0pt}}
                  \parskip 7.25pt
                  \normalbaselines}
\def\pr{\prime}
\def\be{\begin{equation}}
\def\lan{\left\langle}
\def\ran{\right\rangle}
\def\ee{\end{equation}}
\def\barr{\begin{array}}
\def\earr{\end{array}}

\def\nn8{\nonumber\\[10pt]}
\def\l{\left}
\def\r{\right}
\def\dis{\displaystyle}
\def\ed{\end{document}}

\def\ce{{\cal E}}

\begin{document}
\middlespace

\begin{center}
{\bf Breit-Wigner to Gaussian transition in strength functions}

\vskip 0.25Cm

V.K.B. Kota$^{a}$ and R. Sahu$^{a,b}$ \\
$^{a}${\it Physical Research Laboratory, Ahmedabad \,\,380 009, India} \\
$^{b}${\it Physics Department, Berhampur University,
Berhampur\,\, 760 007, India} 
\end{center}
\vskip 0.35Cm
\noindent {\bf ABSTRACT:} 
Employing hamiltonians defined by two-body embedded Gaussian
orthogonal ensemble of random matrices(EGOE(2)) plus a
mean-field producing one-body part, strength functions (for
states defined by the one-body part) are constructed for various
values of the strength of the chaos generating two-body part.
Numerical calculations for six and seven fermion systems clearly
demonstrate Breit-Wigner to Gaussian transition, in the chaotic
domain, in strength functions as found earlier in nuclear shell
model and Lipkin-Meshkov-Glick model calculations.

\vskip 2cm
\noindent PACS number(s): 21.10.Pc, 24.60.Lz, 21.60.Cs 

\newpage

Strength functions (also called local density of states (LDOS)
in literature) of simple modes are basic ingredients of many
particle systems such as atomic nuclei  \cite{Bo-69}.  In the
last few years, with developments in quantum chaos
\cite{Gu-98}, nature of strength functions for finite isolated
interacting many particle systems are being investigated using a
variety of models: (i) Zelevinsky et al using nuclear shell
model \cite{Ze-96};  (ii) Wang et al using the three-orbital
Lipkin-Meshkov-Glick model \cite{Wa-98}; (iii) Borgonovi et al
using a symmetrized coupled two-rotor model
\cite{Bo-98}; (iv) Benet et al using a chaotic model of two
coupled quartic oscillators \cite{Se-00}. Results from the shell
model and Lipkin-Meshkov-Glick model studies clearly showed that
strength functions exhibit, for interacting particle systems,
Breit-Wigner to Gaussian transition. Moreover it is seen that
the transition takes place much after the onset of chaos in
energy levels (i.e. much after level fluctuations start
following GOE). In order to establish that this transition is
generic to interacting particle systems, we carried out one plus
two-body embedded Gaussian orthogonal ensemble (EGOE(1+2))
calculations and the results are reported in this short
communication (see also \cite{Fl-97,Ko-00}).

Given a compound state $\phi_k$, the probability of its decay
into stationary states $\psi_E$ (generated by $H$) is given by
$\l|\lan \phi_k \mid \psi_E\ran\r|^2$. Then the strength
function $F_k(E)$ is,
\be
\l|\l.\phi_k\ran\r. = \dis\sum_E\;C^E_k\;\l|\l.\psi_E\ran
\r.\;\;\Rightarrow\;\;
F_k(E) = \dis\sum_{E^\pr}\; \l|C^{E^\pr}_k\r|^2\;\delta(E-E^\pr) = \lan
\delta(H-E) \ran^k
\ee
The standard form \cite{Bo-69}, normally employed in many
applications in nuclear physics, for strength functions is the
Breit-Wigner (BW) form characterized by a spreading width
$\Gamma_k$,
\be
F_{k:BW}(E) =
\dis\frac{1}{2\pi}\;\dis\frac{\Gamma_k}{(E-{\overline{E_k}})^2 +
\Gamma^2_k/4}
\ee
where ${\overline{E_k}} = \lan k \mid H \mid k \ran =
\int^\infty_{-\infty} F_k(E) E \;dE$. Starting with $H=h(1) +
\lambda V(2)$ (a one plus two-body interaction as in the shell model)
with $h(1)$ defining the $\phi_k$ states, it is easily seen that
the assumptions that give $F_{k:BW}(E)$ will break down when the
mixing is strong. Thus it is expected that the form of
$F_k(E)$ will be different from BW for large $\lambda$. In order
to find the form of strength functions in large $\lambda$ limit,
it is useful to think of $\phi_k$ as a compound state generated
by the action of a transition operator ${\cal O}$ on a state
$\psi_{E_k}$ (for example ground state),
\be
\barr{c}
\l|\l.\phi_{E_k}\ran\r. = \dis\frac{{\cal O}\;\l|\l.\psi_{E_k}\ran\r.}{
\l[\lan\psi_{E_k} \mid {\cal O}^\dagger {\cal O} \mid \psi_{E_k}
\ran\r]^{1/2}}\;\;;\;\;\;\lan \phi_{E_k}\mid\phi_{E_k}\ran = 1 \nn8
\l|\l.\phi_{E_k}\ran\r. =
\dis\sum_{E}\;C^E_{E_k}\;\l|\l.\psi_E\ran\r. \;\;\;\Rightarrow \;\;\;
F_{k;{\cal O}}(E) = \dis\frac{\lan\lan {\cal O}^\dagger \delta(H-E)
{\cal O} \delta(H-E_k) \ran\ran}{\lan\lan {\cal O}^\dagger {\cal
O} \delta(H-E_k) \ran\ran} 
\earr
\ee
The numerator in the expression for $F_k(E)$ in (3) is nothing
but the bivariate strength density generated by the operator
${\cal O}$ \cite{Fr-88} and then it is easy to recognize that
$F_k(E)$ is the conditional density of this bivariate strength
density. It is known \cite{Fr-88} that EGOE($k$), the embedded
GOE of $k$-body interactions (see Ref. \cite{Br-81} for the
definition of EGOE(k)), in general gives bivariate Gaussian
strength densities. Therefore, for EGOE($2$), which models a
generic two-body hamiltonian, the strength functions take
Gaussian form (conditional density of a bivariate Gaussian is a
Gaussian). However, for generic interacting particle systems, it
is more appropriate to consider one plus two-body EGOE(1+2)
hamiltonians $\{H\}=h(1) + \lambda \{V(2)\}$ where $h(1)$ is the
one-body mean-field producing part and $\{V(2)\}$ is the chaos
generating two-body part; note that $\{V(2)\}$ is GOE in
two-particle space with unit matrix elements variance and
$\lambda$ is the interaction strength. Starting with EGOE(1+2)
it is to be expected that for sufficiently large values of
$\lambda$, EGOE(2) description should be valid and therefore (3)
gives the shape of the strength function to be Gaussian for
large $\lambda$; this result is indeed seen in numerical
calculations. Now the important questions are: (i) how $F_k(E)$
changes as $\lambda$ is varied and (ii) from which value of
$\lambda$ strength functions take Gaussian form. 

For further understanding of the nature of $F_k(E)$, we performed
EGOE(1+2) calculations in 924 dimensional space generated by six
fermions ($m=6$) in twelve single particle states ($N=12$) and
similarly in the 3432 dimensional $N=14$ and $m=7$ space.  The
single particle energies employed are $\epsilon_i=i+(1/i)$,
$i=1,2,\ldots,N$ as in \cite{Fl-96}. For various values of
$\lambda$ in $h(1)+\lambda \{V(2)\}$, strength functions are
constructed by choosing the $\phi_k$ states to be the mean-field
$h(1)$ states defined by the distribution of $m$ particles in
the $N$ single particle states; their energies $\ce_k$ are
$\ce_k=\lan k | h(1) + \lambda \, V(2) | k \ran$. In the
calculations $E$ and $\ce_k$ are zero centered for each member
and scaled by the spectrum ($E$'s) width $\sigma$; ${\hat{E}}_k
= (\ce_k-\epsilon)/\sigma$ and ${\hat{E}} =
(E-\epsilon)/\sigma$.  For each member $|C_k^E |^2$ are summed
over the basis states $\l.\l| k \r. \ran$ in the energy window
${\hat{E}}_k \pm \Delta$ and then ensemble averaged
$F_{{\hat{E}}_k} ({\hat{E}})$ vs ${\hat{E}}$ is constructed as a
histogram; the value of $\Delta$ is chosen to be 0.025 for
$\lambda \leq 0.1$ and beyond this $\Delta=0.1$.  Results for
${\hat{E}}_k =0$ are shown for $\lambda=$ 0.05, 0.1, 0.15, 0.2,
0.3 and 0.5 in Fig. 1; in the plots $\int F_{\ce_k}({\hat{E}})\;
d{\hat{E}} = 1$.  From the figures it is clearly seen that there
is BW to Gaussian transition in $F_k(E)$. A measure for this
transition is,
\be
R(\lambda)=\dis\frac{\dis\sum_i \l[F_k^{(\lambda)}(E_i)
-F_{k:BW}(E_i)\r]^2}{
\dis\sum_i \l[F_{(k:BW)}(E_i) -F_{k:ED}(E_i)\r]^2}
\ee
In (4) $E_i$ are defined by the center of each bin in the
histogram representing $F_k^{(\lambda)}(E)$ and $ED$ corresponds
to Gaussian with Edgeworth corrections; the ED incorporates
\cite{Ke-69} skewness ($\gamma_1$) and excess ($\gamma_2$)
corrections. Strength function $F_k^{(\lambda)}(E)$ is BW for
$R=0$ and Gaussian (or ED) for $R=1$. The value of the
interpolating parameter $\lambda= \lambda_{F_k}$ for onset of
the transition from BW to Gaussian is taken to be
$R(\lambda_{F_k})=0.7$. For $\lambda=$ 0.05, 0.1, 0.15, 0.2,
0.3, 0.5 the values of $R$ are 0.13, 0.06, 0.21, 0.7, 0.9, 1.03
for the six particle example. For the seven particle example the
corresponding numbers are 0.15, 0.02, 0.45, 0.75, 0.93, 1.03
respectively. Thus, with the measure $R$ in (4), $\lambda_{F_k}
\approx 0.2$ for both the 6 and 7 particle cases.

In Fig. 2 shown are EGOE(1+2) results for various values of
$\lambda$ for the nearest neighbour spacing dustribution and
$\overline{\Delta_3}(L)$, $0 \leq L \leq 40$ statistic. The
$\lambda_{F_k}$ deduced from the results in Fig. 1 should be
compared with $\lambda = \lambda_c \sim 0.08$ and 0.06,
derived via Fig. 2, for six and seven particle examples
repectively for onset of chaos in level fluctuations. The
$\lambda_c$ numbers are consistant with $(N m^2)^{-1}$ variation
as expected from the theory given in \cite{Ja-97}. It should be
noted that in the $(2s,1d)^{m=12,J^\pi T=0^+ 0}$ shell model
example considered in \cite{Ze-96}, $\lambda_c \sim 0.3$ and
$\lambda_{F_k} \sim 0.6$.  Thus the BW form for $F_k(E)$, which
begins some what before $\lambda$ approches $\lambda_c$ (for
$\lambda << \lambda_c$ the strength functions are delta
functions with perturbative corrections), extends much into the
chaotic domain (defined by $\lambda > \lambda_c$) and the
transition to Gaussian shape takes place in the second layer
defined by $\lambda_{F_k}$ ($\lambda_{F_k} >> \lambda_c$). As
the $\lambda_{F_k}$ in our two examples did not show much
variation with $(N,m)$, it appears that $\lambda_{F_k}$ will
have weak dependence on $m$ unlike $\lambda_c$. Perhaps for
sufficiently large $m$, the transition to Gaussian form takes
place in the thermalization regime discussed by Flambaum and
Izrailev \cite{Fl-97} (see also \cite{Ho-95}).  It should be
remarked that EGOE($2$) gives Gaussian form for state densities
for sufficiently large $m$ and its extension to partial state
densities (which are sums of strength functions) is indeed the
basis of statistical nuclear spectrocopy \cite{Fr-88,Fr-82}.

In conclusion, the results of the present EGOE(1+2) study (Figs.
1,2) and the earlier shell model \cite{Ze-96} and
Lipkin-Meshkov-Glick model \cite{Wa-98} analysis establish
firmly the Breit-Wigner to Gaussian transition for strength
functions, in the chaotic domain, in interacting particle
systems such as atomic nuclei.

---------------

\vskip 0.2cm
\noindent {\bf Acknowledgements}
\vskip 0.1cm
This work has been partially supported by DST(India).

\newpage

\baselineskip=22.5pt
\begin{center}
{\bf FIGURE CAPTIONS}
\end{center}

\noindent {\bf Fig. 1} Strength functions for EGOE(1+2) for
various values of the interaction strength $\lambda$: (i)  for a
system of 6 fermions in 12 single particle states with 25
members; (ii) for a system of 7 fermions in 14 single
particle states (due to computational constraints, here only one
member is considered just as in \cite{Fl-96}).  In the figure,
the histograms are EGOE(1+2) results and continuous curves are
BW fit.  For the 6 fermions case, the dotted curves are Gaussian
for $\lambda \leq 0.15$ and Edgeworth corrected Gaussians (ED)
for $\lambda > 0.15$. Similarly for the 7 fermions case, the
dotted curves are Gaussian for $\lambda \leq 0.1$ and Edgeworth
corrected Gaussians (ED) for $\lambda > 0.1$. See text for
further details.

\noindent {\bf Fig. 2} Nearest neighbour spacing distribution
$P(S)$ vs $S$ and Dyson-Mehta $\overline{\Delta_3}(L)$ statistic
for $0 \leq L \leq 40$ for various values of the interaction
strength $\lambda$ in EGOE(1+2). For $P(S)$, histograms are
EGOE(1+2) results, dashed curves are Poisson and continuous
curves are Wigner distribution. For $\overline{\Delta_3}(L)$,
filled circles are EGOE(1+2) results, dashed curves are Poisson
and continuous curves are for GOE. Results are shown for the six
and seven particle examples considered in Fig. 1.

\ed
\vskip 0.2cm

{\small
\baselineskip=14pt
\begin{thebibliography}{99}

\bibitem{Bo-69} A. Bohr and B. Mottelson, {\it Nuclear
Structure}, Volume 1 (Benjamin, New York, 1969).

\bibitem{Gu-98} T. Guhr, A. M\"{u}ller-Groeling, and H.A.
Weidenm\"{u}ller, Phys. Rep. {\bf 299}, 189 (1998).

\bibitem{Ze-96} V. Zelevinsky, B.A. Brown, N. Frazier, and M.
Horoi, Phys. Rep. {\bf 276}, 85 (1996); N. Frazier, B.A. Brown,
and V. Zelevinsky, Phys.  Rev. C {\bf 54}, 1665 (1996).

\bibitem{Wa-98} W. Wang, F.M. Izrailev, and G. Casati, Phys. Rev.
E {\bf 57}, 323 (1998).

\bibitem{Bo-98} F. Borgonovi, I.  Guarneri, and F.M. Izrailev,
Phys. Rev. E {\bf 57}, 5291 (1998).

\bibitem{Se-00} L. Benet, F.M. Izrailev, T.H. Seligman, and A.
Suarez-Moreno, chao-dyn/9912035 (2000).

\bibitem{Fl-97} V.V. Flambaum and F.M. Izrailev, Phys. Rev. E
{\bf 56}, 5144 (1997).

\bibitem{Ko-00} V.K.B. Kota and R. Sahu, nlin-CD/0006003 (2000);
Phys. Rev. E (in press); V.K.B. Kota, R. Sahu, K.Kar, J.M.G.
G\'omez, and J. Retamosa, nucl-th/0005066 (2000).

\bibitem{Fr-88} J.B. French, V.K.B. Kota, A. Pandey, and S.
Tomsovic, Ann. Phys. (N.Y.) {\bf 181}, 235 (1988).

\bibitem{Br-81} T.A. Brody, J. Flores, J.B. French, P.A. Mello,
A. Pandey, and S.S.M. Wong, Rev. Mod. Phys. {\bf 53}, 385 (1981).

\bibitem{Fl-96} V.V. Flambaum, G.F. Gribakin, and F.M. Izrailev,
Phys. Rev. E {\bf 53}, 5729 (1996).

\bibitem{Ke-69} M.G. Kendall and A. Stuart, {\it Advanced Theory
of Statistics}, Volume 1, 3rd edition (Hafner Publishing
Company, New York, 1969).

\bibitem{Ja-97} Ph. Jacquod and D.L. Shepelyansky, Phys. Rev.
Lett. {\bf 79}, 1837 (1997).

\bibitem{Ho-95} M. Horoi, V. Zelevinsky, and B.A. Brown, Phys.
Rev. Lett. {\bf 74}, 5194 (1995).

\bibitem{Fr-82} J.B. French and V.K.B. Kota, Ann. Rev.  Nucl.
Part. Sci. {\bf 32}, 35 (1982), S.S.M. Wong, {\it Nuclear
Statistical Spectroscopy} (Oxford University Press, New York,
1986); V.K.B. Kota and K. Kar, Pramana - J.  Phys. {\bf 32}, 647
(1989).

\end{thebibliography}
}
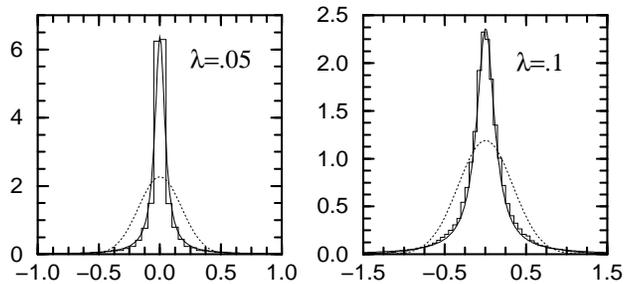
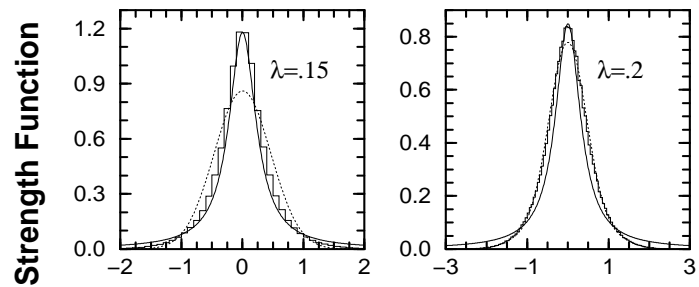
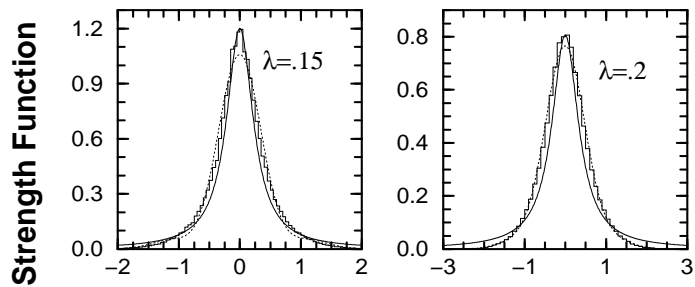
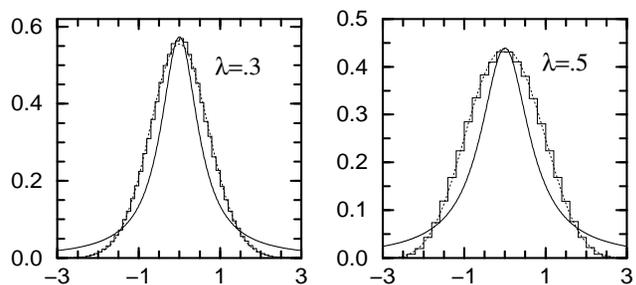
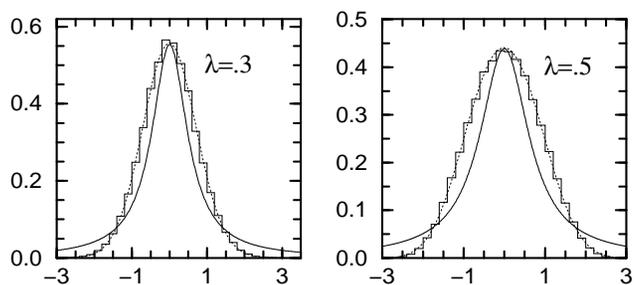
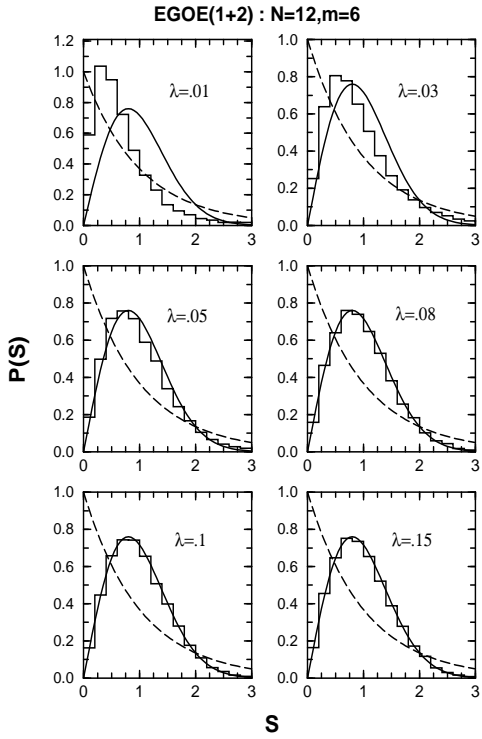
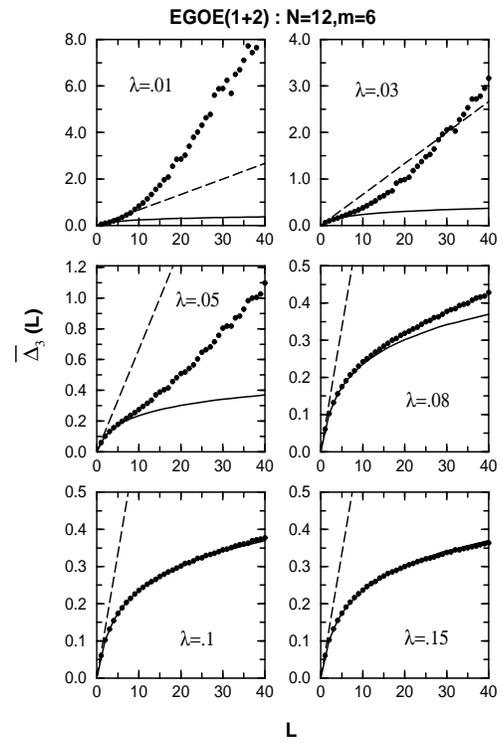
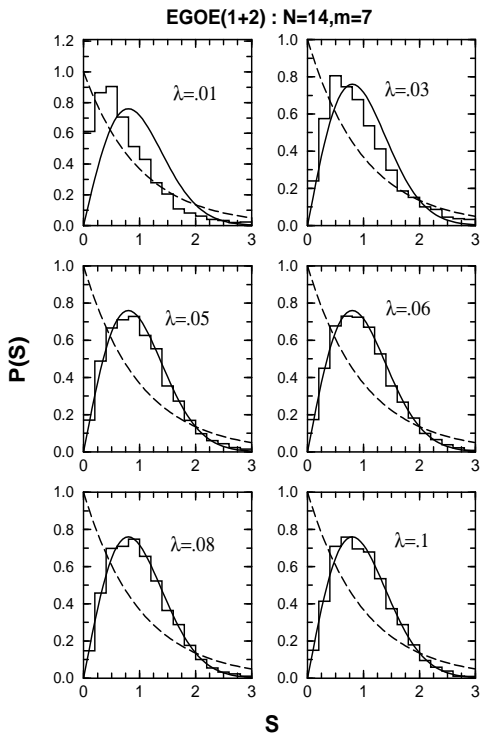
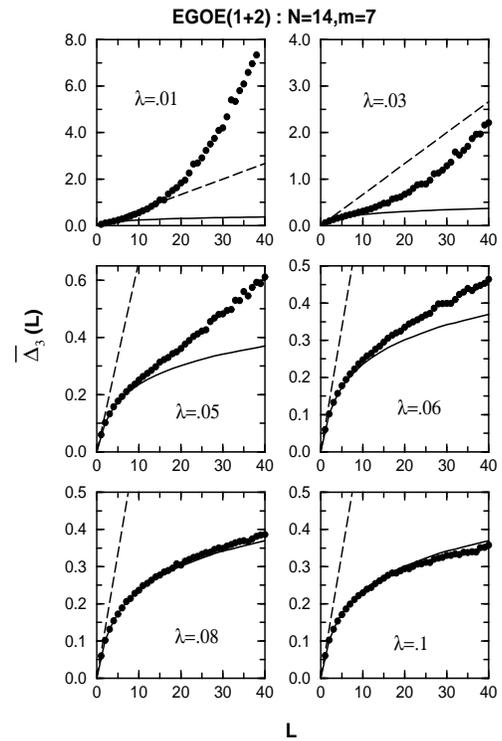
